\documentclass[conference]{IEEEtran}
\IEEEoverridecommandlockouts
\usepackage{cite}
\usepackage{amsmath,amssymb,amsfonts}
\usepackage{algorithmic}
\usepackage{graphicx}
\usepackage{textcomp}
\usepackage{xcolor}
\usepackage{comment}
\usepackage{acronym}

\def\BibTeX{{\rm B\kern-.05em{\sc i\kern-.025em b}\kern-.08em
    T\kern-.1667em\lower.7ex\hbox{E}\kern-.125emX}}

\begin{document}

\newacro{TX}{transmitter}
\newacro{PLL}{phase-locked loop}
\newacro{IoT}{Internet-of-Things}
\newacro{EVM}{error vector magnitude}
\newacro{PSK}{phase-shift keying}
\newacro{DPSK}{differential phase-shift keying}
\newacro{RMS}{root-mean-square}
\newacro{TP}{trigger pulse}
\newacro{ISF}{impulse sensitivity function}
\newacro{BBP}{baseband pulse}

\title{236~$\mu$W Direct-RF PLL-Free Multi-PSK Transmitter Using Oscillator-Based Phase Synthesis
}

\author{
\IEEEauthorblockN{Meysam Sohani Darban$^{1}$, Fariborz Lohrabi Pour$^{2}$, Dong Sam Ha$^{1}$, Jeffrey Sean Walling$^{1}$}
\IEEEauthorblockA{$^{1}$Virginia Polytechnic Institute and State University, Blacksburg, VA, USA\\
$^{2}$University of North Carolina at Charlotte, Charlotte, NC, USA\\
Emails: meysam97@vt.edu, flohrabi@charlotte.edu, ha@vt.edu, jswalling@vt.edu}
}

\maketitle

\begin{abstract}
This paper presents a compact, low-power, direct RF multi-\ac{PSK} \acf{TX} that eliminates the need for a \acf{PLL} by performing phase modulation directly within a ring oscillator. The proposed architecture exploits synchronized charge extraction at the oscillator’s transition points to induce controlled phase shifts while maintaining constant amplitude and frequency. A time-domain multi-triggering technique is introduced to enable reconfigurable multi-mode modulation, supporting 16-\ac{PSK}, 8-\ac{PSK}, Q\ac{PSK}, and B\ac{PSK} within a unified hardware structure. The \ac{TX} circuit is fabricated in a 22-nm FD-SOI process and operates in the ISM band at 2.4~GHz. Measurement results indicate a symbol rate of 2~MSps with a maximum \acf{EVM} of 5.13\%~rms. The core \ac{TX} occupies $23 \times 17.6~\mu\text{m}^2$ and consumes 236~$\mu$W, excluding the output driver, which delivers $-$10~dBm output power over a 60~MHz bandwidth. The proposed design achieves a favorable trade-off between power consumption, circuit complexity, and modulation flexibility, making it well-suited for low-power wireless applications.
\end{abstract}



\section{Introduction}\label{intro}

The rapid proliferation of ultra-low-power wireless systems for \acf{IoT}, biomedical, and distributed sensing has exposed fundamental limitations in conventional RF \acf{TX} architectures~\cite{ref_PLL_TX_2,ref_table_4}. Although \acf{PLL}-based \acp{TX} dominate in modern systems, their power consumption, complexity, and die area make them prohibitive for energy-constrained platforms operating at sub-milliwatt power levels while maintaining acceptable modulation accuracy~\cite{ref_table_1,ref_table_3}. This underscores the need for alternative architectures that bypass the overhead of frequency synthesis while still supporting robust and reliable modulation.

\Acf{PSK} remains a preferred modulation scheme for low-power systems due to its resilience to amplitude noise and channel fading~\cite{ref_iot}. However, most existing \ac{PSK} \acp{TX} rely on \ac{PLL}s or discrete phase-synthesis techniques (e.g., controlled delay elements), leading to phase mismatch, linearity degradation, calibration overhead, and large area consumption~\cite{ref_table_1, ref_table_6}, while failing to exploit the oscillator's intrinsic phase dynamics, and thereby limiting their efficiency in low-power operation.

This work demonstrates that scalable \ac{PSK} can be achieved directly via oscillator phase dynamics, without reliance on a \ac{PLL} or a phase-interpolation network. Despite prior oscillator-based approaches that suffered from limited controllability or coarse phase resolution, the proposed architecture introduces a deterministic phase manipulation based on synchronized charge extraction at the oscillator’s maximum phase-sensitivity points~\cite{ref_pn}. Leveraging the oscillator’s natural charge-to-phase translation mechanism, repeatable phase shifts are achieved while maintaining constant amplitude and frequency.

A key innovation of this work is the introduction of a multi-triggering modulation strategy that redefines how phase resolution is achieved in RF \acp{TX}. The proposed technique encodes baseband information by accumulating discrete phase adjustments of a constant value in the time domain, rather than using analog tuning elements (e.g., large capacitor banks~\cite{ref_pre}), thereby enabling multi-level \ac{PSK} with a single compact hardware structure. This removes the trade-off between phase resolution and complexity, while supporting reconfigurable modulation without additional hardware overhead. The resulting architecture represents a transition toward digitally assisted, oscillator-centric RF \acp{TX}, where phase modulation is employed directly in the signal generation process.

This article is organized as follows. Section~\ref{Design} presents the theoretical foundation of oscillator-based phase modulation and describes the proposed multi-\ac{PSK} \ac{TX} architecture and performance considerations. Section~\ref{Results} describes the measurement results, and finally, Section~\ref{con} concludes the paper.

\section{Direct RF Multi-PSK Transmitter}\label{Design}

This section explains the concept of the oscillator-based \ac{PSK} modulation, providing a description of the proposed \ac{TX} architecture with analysis of its performance considerations.

\subsection{Oscillator-Based Phase Shift Keying Concepts}

\begin{figure}[t]
\centering
\includegraphics[width= 0.85\columnwidth]{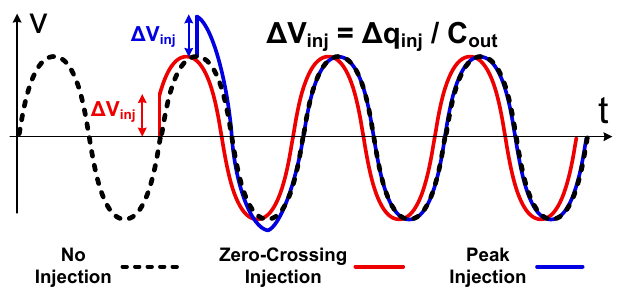}
\caption{Charge injection effect at different moments of an oscillator's output.}
\label{fig_wave}
\end{figure}

In a \ac{PSK} \ac{TX}, the carrier's phase varies while its amplitude remains constant. The oscillator can perform this function via direct modulation, but in most cases, it is done in the baseband. In the case of direct modulation, one way to characterize the oscillator's performance is to view it as charging and discharging its output nodes within each operating period while maintaining stable oscillation. If any additional charge is injected or drawn from these output nodes, it can perturb both the amplitude and phase of the signal. Fig.~\ref{fig_wave} illustrates the effect of injecting the charge $\Delta q_{inj}$ into the output capacitor of the oscillator ($C_{out}$) at two critical time instances. The red line shows the signal deviation when injecting at the zero-crossing point, while the blue line is the response of the oscillator to the charge injection at the peak amplitude. This extra charge causes a voltage change across the output capacitor equal to:

\begin{equation}
   \Delta V_{inj} = \dfrac{\Delta q_{inj}}{C_{out}}.
    \label{eq_charg}
\end{equation}

Owing to the self-limiting nature of the oscillator and the constraints of the power supply, any variations in amplitude are suppressed, which ensures that the output remains constant \cite{ref_ring}. Alternatively, the effect of the excess charge on the phase of the voltage waveform remains. This phase variation depends on the injection time. As discussed in \cite{ref_pn}, the \acf{ISF} of an oscillator relates this dependency. For a ring oscillator, if the excess charge is injected at the signal's amplitude peaks, there will be no phase shift in the waveform. Conversely, the maximum phase shift occurs when the charge is injected at the zero-crossing points.

\begin{figure}[t]
\centering
\includegraphics[width= 0.85\columnwidth]{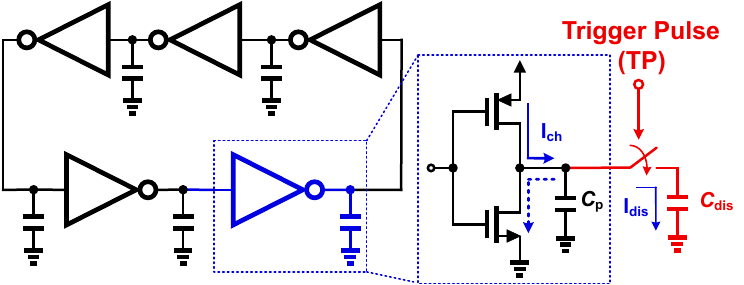}
\caption{Concept of charge extraction on a ring oscillator.}
\label{fig_osc}
\vspace{-5mm}
\end{figure}

The ring oscillator shown in Fig.~\ref{fig_osc} consists of an odd number of inverters with a charging current of $I_{ch}$ followed by a total output capacitor (parasitic and fixed) of $C_{p}$. While oscillating, the oscillator charges/discharges $C_{p}$ during each cycle. The output current of each stage and the amount of charge that sinks or sources each node's equivalent capacitor determine the circuit's frequency. If the capacitor $C_{dis}$, with no initial charge, is momentarily connected to an oscillator output node via an ideal switch controlled by a \ac{TP}, a portion of $I_{ch}$ flows into $C_{dis}$ as a discharge current ($I_{dis}$). This process extracts the charge that would otherwise contribute to $C_{p}$. If the switch turns on for a short period, this acts as the charge injection described in Fig.~\ref{fig_wave}.

The \ac{TP}'s timing relative to the output signal defines the window at which charge extraction occurs. Two main factors that determine the extent of phase adjustment achievable at the output are the timing of the \ac{TP}, due to the \ac{ISF}, and the ratio of $C_{dis}$ to $C_{p}$, which dictates the amount of charge that can be drawn from the output node. For a ring oscillator, the ISF peaks at the transition points from 0 to $V_{DD}$ \cite{ref_ring}. By ensuring that the \ac{TP}'s timing is synchronized to this transition point and using a properly sized capacitor at one node of the ring oscillator, the circuit develops controllable phase shifts while maintaining constant frequency and amplitude, which is essential for a \ac{PSK} \ac{TX}.  Controlling the capacitor and the \ac{TP}'s timing leads to a direct RF-oscillator-based \ac{PSK} modulator.

\subsection{Transmitter Circuit Design}

\begin{figure}[t]
\centering
\includegraphics[width= 0.75\columnwidth]{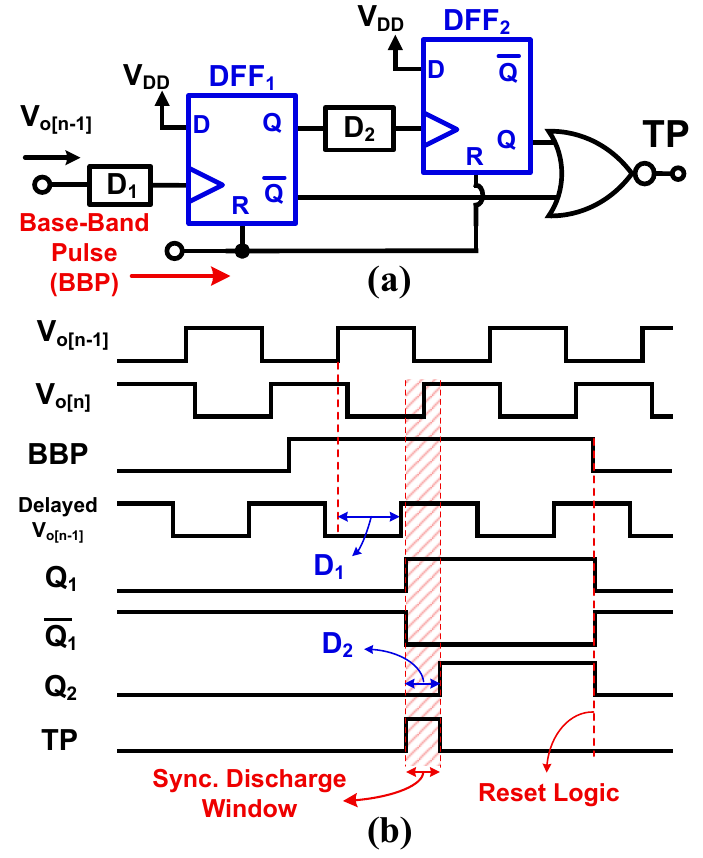}
\caption{(a) Schematic of proposed control unit circuit, and (b) logic wave forms corresponding to creating a \ac{TP} window.}
\label{fig_cu}
\vspace{-5mm}
\end{figure}

To provide a timing window for \ac{TP}, a digital control unit has been designed that generates a window, synchronized to the rising edge of the oscillator's output, for each \ac{BBP} received from the baseband modem. Fig.~\ref{fig_cu}~(a) displays the proposed circuit. The control unit will utilize the output of one stage of oscillator, $V_{o[n-1]}$, and generate a narrow timing window over the rising edge of the next stage's output, $V_{o[n]}$. 

As illustrated in Fig.~\ref{fig_cu}~(b), when \ac{BBP} is set to 0, the DFFs are in the reset state. When \ac{BBP} is 1, the first DFF can function on the rising edge of its clock, provided by the $D_{1}$ delay path. Then, by sensing the rising edge, the $Q_{1}$($\overline{Q_{1}}$) will be set to 1(0). Changing the $\overline{Q_{1}}$ will raise the \ac{TP} to 1, which is the \ac{TP} window's first edge. After the rising edge of $Q_{1}$ passes through the $D_{2}$'s delay line, it changes the second DFF's outputs. As a result, \ac{TP} returns to 0 again, creating the second edge of the \ac{TP} window. Lastly, when the \ac{BBP} returns to 0, the control unit's signals will be reset and ready for the next data transmission cycle. As shown in  Fig.~\ref{fig_cu}~(b), $D_{1}$ determines the position of the window, and $D_{2}$ defines its width. $D_{1}$ and $D_{2}$ were optimized to keep the rising edge within the window across all PVT variations so the charge extraction can trigger the \ac{ISF} most effectively.

\begin{figure}[t]
\centering
\includegraphics[width= 0.86\columnwidth]{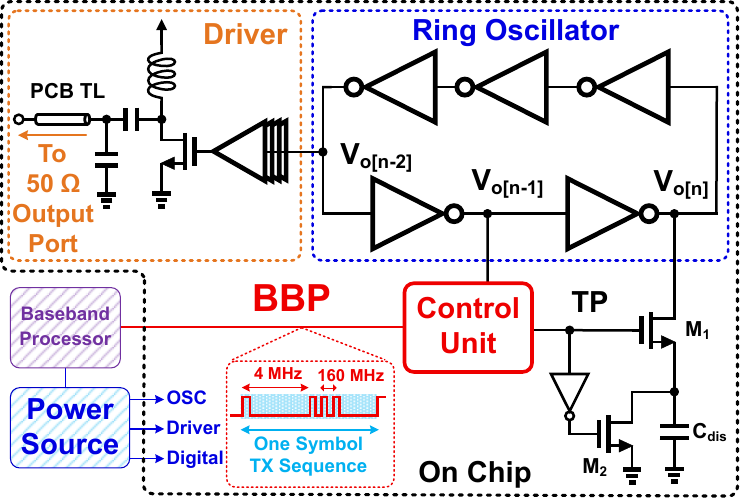}
\caption{Schematic of proposed direct RF \ac{PSK} \ac{TX}.}
\label{fig_tx}
\vspace{-5mm}
\end{figure}

The schematic of the proposed \ac{PSK} \ac{TX} is illustrated in Fig.~\ref{fig_tx}. When the \ac{TP} window turns on the $M_{1}$ switch, it connects the $C_{dis}$ to the $V_{o[n]}$ and draws a portion of its charge while it tries to raise $V_{o[n]}$ to $V_{DD}$, leading to a constant phase shift over the operation phase of the oscillator. In addition, in reset mode, the $M_{2}$ ensures the $C_{dis}$ is in a zero-charge state before connection. The other output of the oscillator is connected to a buffer chain, which drives a power driver transistor, sending the RF data to the output pad via a matching network and the PCB transmission line (TL). A microcontroller can handle the baseband processing and generate the required BBP sequences for the \ac{TX} to send data.

\subsection{Performance Considerations}

The phase shift of this circuit is constant; however, \ac{PSK} requires different phases for the RF carrier. One approach is to change the phase using a bank of capacitors. However, simulation results indicate linearity degradation across different symbols due to non-idealities in the capacitor bank and its switch behavior. Additionally, the capacitor bank increases the die area, which is not fit for a compact all-digital design.

To address this issue, a multi-triggering technique is proposed. As the width of the TP window and the value of $C_{dis}$ are set to provide only $22.5^{\circ}$ of phase adjustment each time the TP window is applied. This phase adjustment is considered the phase-shift least significant bit (LSB) of the \ac{TX}'s modulation, thereby enabling standard 16\ac{PSK}. Since all required phases of the target modulation are integer multiples of the circuit's LSB, to transmit each symbol from this modulation, the baseband must apply an integer number of BBPs in one baseband clock period. The multi-triggering technique and the proposed ring oscillator-based \ac{PSK} result in a simple, compact \ac{PSK} \ac{TX}.

One advantage of the multi-triggering strategy is that the \ac{TX} can dynamically define its effective LSB based on the number of generated triggers. For instance, if the processor generates two BBPs per circuit LSB, the \ac{TX} effectively operates as an 8\ac{PSK} modulator. Consequently, the proposed circuit can support multiple \ac{PSK} modulation modes.

A limitation of the multi-triggering strategy is its inherent hysteresis, which can be solved by employing \acf{DPSK} instead of conventional PSK. In this method, the \ac{TX} initially sends a reference symbol using a single BBP to establish a phase reference at the receiver. Subsequent symbols are generated using a variable number of BBPs, 1-16, to convey the desired data as illustrated in Fig.~\ref{fig_tx}.

The absence of a continuous reference frequency is architecturally justified by the well-established philosophy of asymmetric complexity, which shifts the synchronization burden to the receiver and prioritizes extreme energy efficiency and a minimal die area for the \ac{TX}. With direct time-domain phase modulation of the oscillator via synchronous charge extracting, this design avoids the power-hungry frequency synthesis normally required in RF transmission. Although the lack of a feedback mechanism might lead to frequency drift, the design automatically compensates for this issue because \ac{DPSK} is used as the modulation method. This allows the receiver equipped with Clock and Data Recovery (CDR) to effectively phase-lock onto the carrier and sync with the bit stream, thereby moving the burden of frequency stabilization from the power-limited \ac{TX} to the power-rich receiver. Moreover, if higher-frequency accuracy is needed, a low-duty-cycle PLL may be added on the \ac{TX} side for occasional calibration. Such a burst-mode calibration technique will enable correcting long-term frequency centering without the overhead associated with a fully functional frequency synthesizer.

\section{Measurement Results}\label{Results}

\begin{figure}[t]
\centering
\includegraphics[width= 0.85\columnwidth]{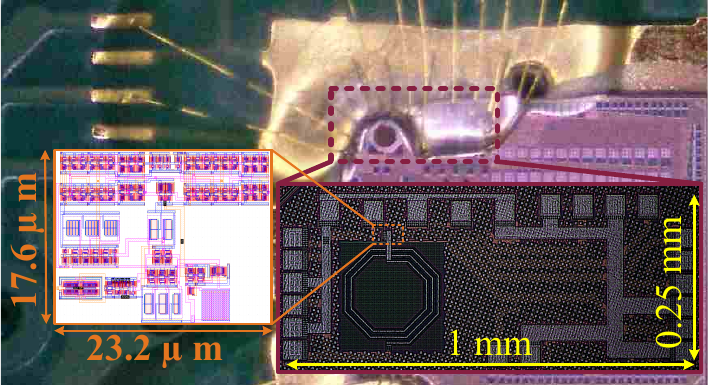}
\caption{Chip photograph and the layout of the proposed \ac{TX}.}
\label{fig_chip}
\end{figure}

\begin{figure}[t]
\centering
\includegraphics[width= 0.87\columnwidth]{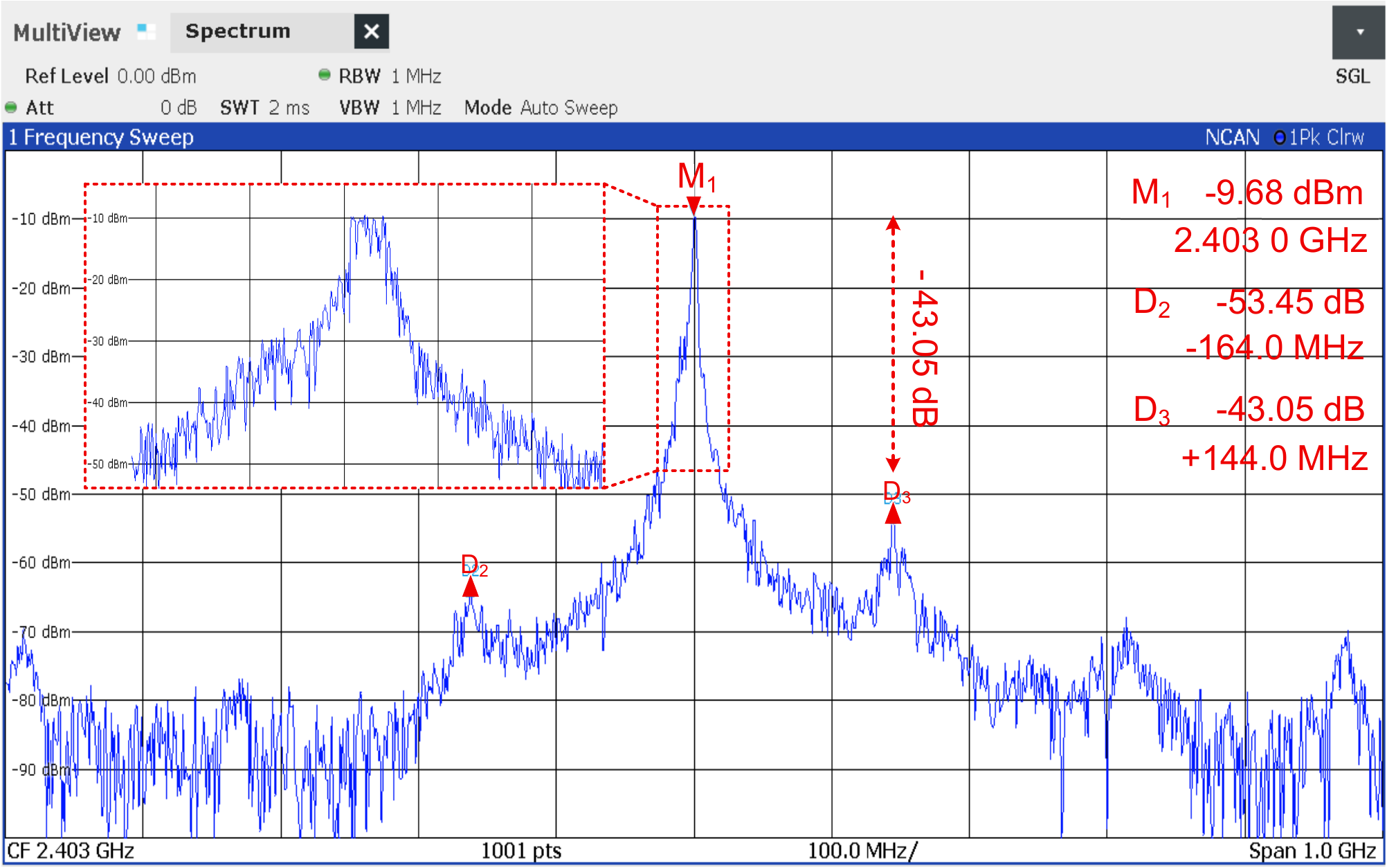}
\caption{Measured output spectrum.}
\label{fig_spectrum}
\vspace{-5mm}
\end{figure}

The proposed 2.4~GHz multi-\ac{PSK} \ac{TX} was fabricated using 22-nm FD-SOI process technology. The die photo and the core digital \ac{TX}'s layout are shown in Fig.~\ref{fig_chip}. The total occupied area (including pads) is $1\times0.25\rm~mm^{2}$, whereas the all digital core \ac{TX}, excluding the driver, only occupied $23.2\times17.6\rm~\mu m^{2}$. The \ac{TX}'s maximum symbol rate is set at 2 MSps due to two primary constraints: the time required for the oscillator to settle into an adjusted phase and the digital processing unit's speed limits in generating TP windows. To achieve this rate using \ac{DPSK}, the design employs a 4~MHz baseband clock for symbols and a 160 MHz BBP window rate.

\begin{figure}[t]
\centering
\includegraphics[width= 0.9\columnwidth]{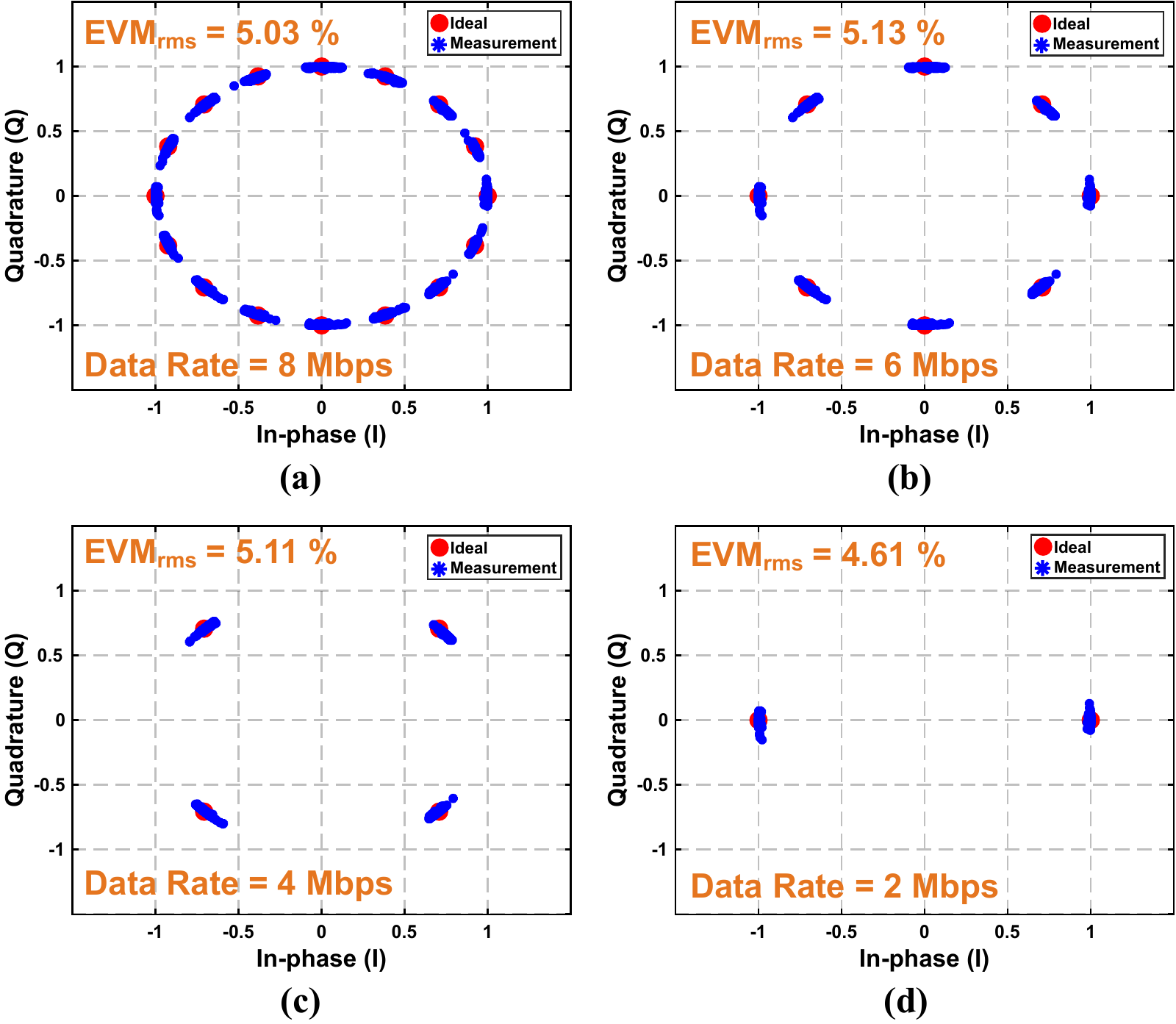}
\caption{\ac{TX}'s constellations for (a) 16\ac{DPSK}, (b) 8\ac{DPSK}, (c) Q\ac{DPSK}, and (d) B\ac{DPSK} at 2~MSps.}
\label{fig_cons}
\vspace{-5mm}
\end{figure}

Fig.~\ref{fig_spectrum} shows the output spectrum of the \ac{TX} with baseband intermodulations on both sides of the main carrier, each at least 43 dB lower than the carrier tone. Fig.~\ref{fig_cons} illustrates the measured \ac{PSK} constellations. Fig.~\ref{fig_cons}~(a) illustrates the constellation for 16\ac{DPSK} vs ideal symbols over 2~MHz, which is equal to 8~Mbps. Based on the measurement results, the \ac{RMS} \acf{EVM} of the circuit is 5.03\%, which is due to the oscillator phase noise. Since the circuit can operate in multi-\ac{DPSK} modes, its performance at the same symbol rate in 8\ac{DPSK}, Q\ac{DPSK}, and B\ac{DPSK} is shown in Figs.~\ref{fig_cons}~(b)--(d), respectively. 

\begin{figure}[t]
\centering
\includegraphics[width= 0.87\columnwidth]{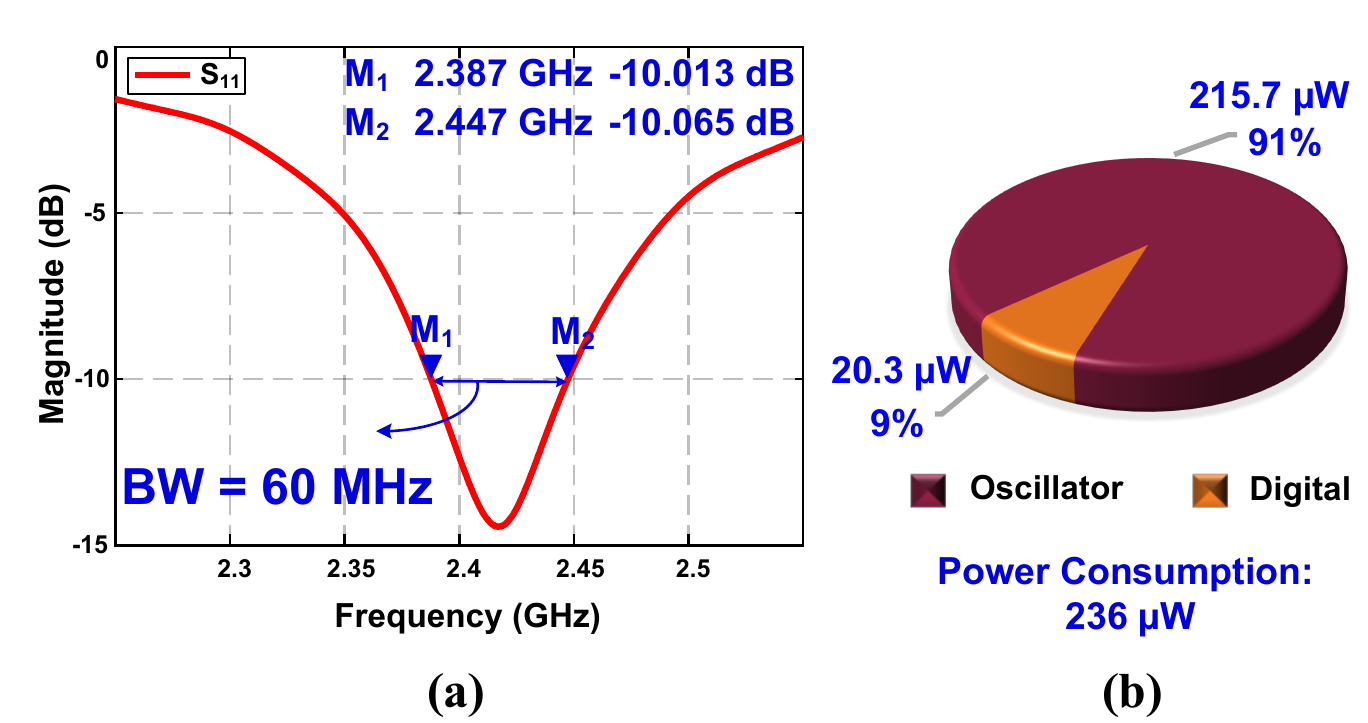}
\caption{(a) Measured output matching performance, and (b) power breakdown.}
\label{fig_s11}
\vspace{-5mm}
\end{figure}

The on-chip matching network, including the PCB interconnection, is measured ($S_{11}$) and shown in Fig.~\ref{fig_s11}~(a), which maintains $S_{11}<$ -10~dB over a 60~MHz bandwidth centered at 2.4~GHz for the \ac{TX}. The \ac{TX}'s power breakdown is displayed in Fig.~\ref{fig_s11}~(b) using measured power and simulation results. The total power consumption, excluding the driver, is 236~$\mu$W. While the output driver's power dissipation is 4.3~mW to deliver -10~dBm output power. Table~\ref{tab_comp} summarizes the design's performance compared to state-of-the-art works, indicating better energy-per-bit efficiency and significantly lower power consumption and core area due to the PLL-free design.

\begin{table}[t]
\caption{Measured Performance Summary and Comparison}
\label{tab_comp}
\resizebox{\columnwidth}{!}{%
\begin{tabular}{|c|c|c|c|c|c|}
\hline
                                                                      & \textbf{This Work}                                                                 & \cite{ref_table_1}                                                     & \cite{ref_table_6}                                        & \cite{ref_table_3}                                     & \cite{ref_table_4}                                   \\ \hline
Technology                                                            & \textbf{\begin{tabular}[c]{@{}c@{}}22 nm\\ FD-SOI\end{tabular}}                    & \begin{tabular}[c]{@{}c@{}}22 nm\\ FD-SOI\end{tabular}                 & 65 nm                                                     & \begin{tabular}[c]{@{}c@{}}22 nm\\ FD-SOI\end{tabular} & 65 nm                                                \\ \hline
\begin{tabular}[c]{@{}c@{}}Standard /\\ Modulation\end{tabular}       & \textbf{\begin{tabular}[c]{@{}c@{}}ISM band/ \\ (16, 8, Q, B)\\ DPSK\end{tabular}} & \begin{tabular}[c]{@{}c@{}}BT, BLE, \\ 802.15.4 / \\ 8PSK\end{tabular} & \begin{tabular}[c]{@{}c@{}}ISM Band /\\ 8PSK\end{tabular} & \begin{tabular}[c]{@{}c@{}}BLE /\\ GFSK\end{tabular}   & \begin{tabular}[c]{@{}c@{}}BLE /\\ GFSK\end{tabular} \\ \hline
Architecture                                                          & \textbf{Direct RF TX}                                                              & PLL-TX                                                                 &Injection Lock PA                                            & PLL-TX                                                 & Polar-TX                                             \\ \hline
\begin{tabular}[c]{@{}c@{}}Data rate (Mbps)\end{tabular}            & \textbf{8 / 6 / 4 / 2}                                                             & 0.125 - 3                                                              & 15                                                        & 1 / 2                                                  & 1                                                    \\ \hline
$EVM_{rms}$ (\%)                                                      & \textbf{4.61 - 5.13}                                                               & 6.07                                                                   & 3.43 / 2.05                                               & -                                                      & -                                                    \\ \hline
\begin{tabular}[c]{@{}c@{}}Energy/bit\\ (nJ/bit)\end{tabular}         & \textbf{0.0295 $^{***}$}                                                             & 0.83                                                                   & 0.57 / 0.48                                               & 1.32 / 0.66                                            & 1.2                                                  \\ \hline
\begin{tabular}[c]{@{}c@{}}$P_{DC}^{  *}$ (mW)\end{tabular}         & \textbf{0.236}                                                                     & 2.49 / 2.97                                                            & 8.48 / 7.2                                                & 1.32                                                   & 1.2                                                  \\ \hline
\begin{tabular}[c]{@{}c@{}}Core Area$^{**}$ (mm$^2$)\end{tabular} & \textbf{408 $\mu m^2$}                                                             & 0.35                                                                   & 0.26                                                      & 0.9                                                    & 0.4                                                  \\ \hline
\end{tabular}%
}

\par\smallskip
\parbox{\columnwidth}{%
\raggedright\scriptsize
$^*$ Excluding driver (PA).
$^{**}$ Estimated TX area.
$^{***}$ 16-DPSK at 8 Mbps.
}
\vspace{-8mm}
\end{table}

\section{Conclusion}\label{con}
This paper presents a low-power, multi-PSK ring-oscillator-based \ac{TX} fabricated in 22-nm FD-SOI technology for the 2.4~GHz ISM band. The architecture eliminates the conventional PLL by performing phase modulation directly within the oscillator via synchronized charge extraction. By employing a time-domain multi-triggering technique, the design supports reconfigurable 16DPSK, 8DPSK, QDPSK, and DBPSK modulation schemes. Measurement results demonstrate a maximum \ac{EVM} of 5.13\% at 2~MSps. Excluding the driver, the core transmitter occupies a compact area of $23.2\times17.6~\mu m^{2}$ and consumes only $236~\mu W$. These results validate the proposed design as a highly energy-efficient and scalable solution for ultra-low-power wireless \ac{IoT} and biomedical applications.


\section{Acknowledgment}
The authors would like to acknowledge GlobalFoundries' University Partner Program for their chip fabrication support.

\bibliographystyle{IEEEtran}
\bibliography{ref}

\end{document}